\documentstyle[aps,prl,twocolumn]{revtex}

\begin{document}
\draft
\title{Breakdown of Luttinger liquid state in one-dimensional frustrated spinless
fermion model}
\author{A. K. Zhuravlev and M. I. Katsnelson}
\address{Institute of Metal Physics, Ekaterinburg 620219, Russia}
\date{\today}
\maketitle

\begin{abstract}
Haldane hypothesis about the universality of Luttinger liquid (LL) behavior
in conducting one-dimensional (1D) fermion systems is checked numerically
for spinless fermion model with next-nearest-neighbor interactions. It is
shown that for large enough interactions the ground state can be gapless
(metallic) due to frustrations but not be LL. The exponents of correlation
functions for this unusual conducting state are found numerically by
finite-size method.
\end{abstract}


One-dimensional (1D) Fermi systems have a number of peculiarities which
distinguish them drastically from 3D ones (for review, see \cite{solyom}).
In particular, gapless (metallic) 1D systems of interacting fermions never
behave as normal Fermi-liquid. Haldane \cite{haldane} have proposed another
class of universality Luttinger liquid (LL) state, namely. It is
characterized by the existence of three branches of low-lying Bose
excitations, density, current, and charge excitations with the
velocities $v_S,v_J$ and $v_N$, correspondingly. The first one is connected with
the variation of the total energy of the system $E$ under the variation
of the total momentum $P,$ $v_S=\delta E/\delta P$; the second one ($v_J$),
with the variation of the energy under the shift of all the particles in
momenta space, that can be done physically by the application of magnetic
flux to the system closed as a ring; and the third one, with the variation of
the chemical potential $\mu$ at the change of the total number of particles
$N,$ $v_N=\left( L/\pi \right) \delta \mu /\delta N,$ with $L$ being the
length of the system. In LL state there are an exact relation between the
velocities
\begin{equation}
\chi \equiv v_Jv_N/v_S^2=1,  \label{chi}
\end{equation}
which is the criterion of LL. The only dimensionless parameter which determines
all the infrared properties of the system (e.g., time and space asymptotics
of fermionic Green functions and susceptibilities) is the ratio
\begin{equation}
e^{-2\varphi }=v_N/v_S=v_S/v_J  \label{lutt2}
\end{equation}
Original arguments by Haldane were based on exact Bethe Ansatz solutions as
well as on the perturbation theory for weakly interacting systems. There are
no general and rigorous proof of this assumption but all the known
analytical and numerical results about 1D fermion systems confirm Haldane
hypothesis \cite{conf,poil}. Anderson \cite{anderson} proposed that some 2D
systems such as copper-oxide superconductors also belong to the class of LL
which made the concept of Luttinger liquid one of the most
``fashionable'' in contemporary many-particle physics. Therefore the
investigation of the status of Haldane hypothesis seems to be of importance.
Here we present a counterexample to this hypothesis basing on exact
numerical results for spinless fermion model.

We proceed with the Hamiltonian
\begin{equation}
H=-t\sum\limits_{i=1}^L\left( c_i^{\dagger }c_{i+1}+c_{i+1}^{\dagger
}c_i\right) +V\sum\limits_{i=1}^Ln_in_{i+1}+V^{\prime
}\sum\limits_{i=1}^Ln_in_{i+2}  \label{hamil}
\end{equation}
where $c_i^{\dagger },c_i$ are Fermi creation and annihilation operators on
site $i,n_i=c_i^{\dagger }c_i.$ Phase diagram of this model has been
investigated by us in \cite{ZKT}. In particular, it has been shown that for
half-occupied case, $\rho =N/L=1/2$ and arbitrarily small $t$ the ground
state turns out to be gapless (metallic) along the line $V=2V^{\prime }.$ It
is the consequence of frustrations which lead to the macroscopically large
degeneracy (finite entropy per site) of the ground state at the Ising limit
$t=0$. A similar result has been obtained also in Ref. \cite{Tsiper}. It is
important that, according to our calculations, the metallic region has
non-zero width in $(V,V^{\prime })$ plane. One can say with certainty that
the gap is zero at $(V/2)-0.6t\le V^{\prime }\le (V/2).$ One can show also
with certainty that the ground state is insulating at $|(V/2)-V^{\prime }|>1$.
To check the Haldane hypothesis we restrict ourselves by the consideration
of the straight line $V=2V^{\prime }$ where the system is definitely
metallic. Similarly, we have the metallic state for $\rho =2/3$ at $%
V^{\prime }=0$ and arbitrarily large $V$ or, vice versa, at $V=0$ and
arbitrarily large $V^{\prime }.$ There are rare examples of metallic state
with strong interactions and it seems to be interesting to check the Haldane
assumptions for this unusual case. As it was already mentioned
the original Haldane hypothesis is based,
on the one hand, on the consideration of exactly integrable systems and, on
the other hand, on the perturbation treatment of systems with weak
correlations. Therefore, its validity in the case under consideration is not
obvious. Note that ``unusual'' character of
metallic state at $\rho =1/2,V\approx 2V^{\prime }$ has been mentioned in
\cite {Tsiper} but without specification what this state is.

We have carried out the calculations of the ground state of the model (\ref
{hamil}) by Lanczos method for finite clusters with the consequent
extrapolation to $L\to\infty $ (for details, see \cite{ZKT}). Velocities of
low-lying excitations has been calculated as \cite{poil,vel}
\begin{eqnarray}
v_S &=&\frac L{2\pi }\left[ E_{1p}(L,N)-E_0(L,N)\right]  \nonumber \\
v_J &=&\frac L{2\pi }\left[ E_a(L,N)-E_0(L,N)\right] \label{veloc} \\
v_N &=&\frac L\pi \left[ E_0\left( L,N+1\right) -2E_0\left( L,N\right)
     +E_0\left( L,N-1\right) \right]     \nonumber
\end{eqnarray}
Here $E_0\left( L,N\right)$ is the ground state energy of the cluster with
$L$ sites for periodic boundary conditions and $N$ particles, $E_a(L,N)$
is ground state energy for antiperiodic boundary conditions (transition to the antiperiodic
conditions corresponds to magnetic flux $\Phi =1/2$ of the flux quantum),
$E_{1p}(L,N)$ is the ground state energy for minimal nonzero total momentum
$P=2\pi /L$.  Then we have verified the criterion of LL $\chi =1$ using
Eq.(\ref{chi}).

The results of the testing calculations for the case $\rho =1/2,V^{\prime
}=0,0\le V<2t$ where the system has to be LL \cite{haldane} are shown in
Fig.1 (open circles and triangles). We also present in the same figure the
calculated values of $\chi$ along the line $V=2V^{\prime }$. One can see
that at $V\le 10t$ we have, within the accuracy of the computations, $\chi
\approx 1$, in an agreement with Haldane hypothesis. However, for $V\ge 30t$
the values of $\chi $ is definitely less than unity, that is obvious even
without extrapolation to $L\rightarrow \infty $ since $\chi (L)<1$ for
finite $L$ and diminishes with $L$ increase. Therefore we have demonstrated
that there are one-dimensional conducting systems of interacting fermions
which {\it are} {\it not} LL. The breakdown of LL picture is caused by the
competitions of nearest-neighbor and next-nearest-neighbor interactions
(i.e. frustrations) which allow the system to be metallic in the limit of
strong interactions. A schematic phase diagram is shown in Fig.2.
The question is still open whether the transition from insulating state
to non-LL conducting state is the direct one or there exists intermediate
conducting LL phase. At the same time, our calculations demonstrate
that for $\rho =2/3$ the relation (\ref{chi}) takes place with the accuracy
of calculations for any values of parameters under consideration even along
the lines $V=0$ or $V^{\prime }=0$. It would be very interesting to
understand analytically the reason for the difference between these two cases
with strong frustrations.

We also have calculated the static correlation functions
\begin{equation}
\label{corr}
\begin{array}{ll}
G\left( R\right) &=\langle c_R^{\dagger}c_0 \rangle  \\ [\bigskipamount]
K\left( R\right) &=\langle \delta n_R\delta n_0 \rangle \\
\end{array}
\end{equation}
where brackets mean the averaging over the ground state, $\delta
n_i=n_i-\rho$. In LL the following asymptotics have to be valid at $R\gg 1$
\cite{haldane}

\begin{equation}
\label{asym}
\begin{array}{ll}
G\left( R\right)  &\sim \sum\limits_{m=0}^\infty C_m\sin \left[ \left(
2m+1\right) k_FR\right] R^{-\eta_m} \\ [\bigskipamount]
K\left( R\right)  &\sim \sum\limits_{m=0}^\infty D_m\cos \left(
2mk_FR\right) R^{-\theta_m}
\end{array}
\end{equation}
where $\eta _m=\frac 12e^{-2\varphi }+2\left( m+\frac 12\right)
^2e^{2\varphi },\theta _m=2m^2e^{2\varphi }$ $\left( m>0\right) ,k_F=\rho /2$
is the Fermi momentum. The most important exponent $\alpha $ determins the
behavior of one-particle distribution function $n\left( k\right) $ near the
Fermi surface
\begin{equation}
n\left( k\right) \approx n\left( k_F\right) -C\textrm{sign}\left( k-k_F\right)
\left| k-k_F\right| ^\alpha , \label{fermi}
\end{equation}
where $\alpha=\eta_0-1.$

However we cannot use these expressions {\it a priori} since for the model
under consideration the system is not always LL. We have found the
asymptotics of the correlation functions by direct computation. It is known
(see, e.g., \cite{Gagliano}) that it is very difficult to find the
correlation exponents from the calculations for a given $L$, even as large
as $L=32.$ Therefore we use the finite size scaling technique \cite{Koma}.
Specifically we use the following procedure.

Our aim is to find the function $\varphi (R)\equiv \langle \phi (0)\phi
(R)\rangle _\infty $ for the infinite chain. Direct calculations give us the
functions $f(R,L)\equiv \langle \phi (0)\phi (R)\rangle _L$ for $R<L$. From
the symmetry considerations we have $f(R,L)=f(L-R,L)$. Let us introduce the
function $r(R,L)$ to have, by definition, $\varphi [r(R,L)]=f(R,L)$.
Therefore
\begin{equation}
\lim_{L\to \infty }r(R,L)=R.  \label{rRL}
\end{equation}
Then we introduce the new variable $\lambda \equiv R/L$ so that
$r(R,L)=L\cdot r^{\prime }(\lambda ,L)$ where $r^{\prime }(\lambda ,L)$ is a
new unknown function. To provide (\ref{rRL}) we have $\lim_{L\to \infty
}r^{\prime }(\lambda ,L)=\lambda $. Also the function $r^{\prime }$
satisfies the condition $r^{\prime }(\lambda ,L)=r^{\prime }(1-\lambda ,L)$.
For small $\lambda $ one has  $r^{\prime }(\lambda )\approx \lambda .$ To
satisfy all these requirements we try the function $r^{\prime }$ as a
Fourier series

\begin{equation}
r^{\prime }(\lambda )=\frac{\sin (\pi \lambda )+a_3\sin (3\pi \lambda
)+a_5\sin (5\pi \lambda )+\dots }{\pi (1+a_3+a_5+\dots )}  \label{Four}
\end{equation}
Using the asymptotic expression similar to (\ref{asym}) for the dependence
$f(R,L)=\varphi \left( Lr^{\prime }(\lambda )\right) $at finite $L$ and
optimizing the result with respect to both $a_n$ and the correlation
exponents we can find the latter with high enough accuracy. At least the
results for the exponents appeared to be accurate enough for the
clusters with $14\leq L\leq 26$ used in our calculations. For the testing
case $V^{\prime }=0, 0<V<2t$ where the system is definitely LL the results
for the correlation exponents coincide with that from Haldane formula (\ref
{asym}) with the accuracy of 0.5\% for the function $G\left( R\right)$ and
8\% for the function $K\left( R\right)$.

In the most interesting case $\chi \ne 1$ we cannot use the expression (\ref
{asym}) and have to restrict ourselves only by the consideration of the
leading terms in the asymptotics of correlation functions which are tried in
the following form

\begin{equation}
\label{asym2}
\begin{array}{ll}
G(R) &\sim (C_1+C_2\sin (\frac \pi 2R))/R^\gamma \\ [\bigskipamount]
K(R) &\sim (D_1+D_2(-1)^R)/R^\delta   \\
\end{array}
\end{equation}
(we consider the case $\rho =1/2$). To diminish the number of states in the
Gilbert space under consideration we use only the states which
has the same (minimal) energy for  $V=2V^{\prime}$ and $t=0$ which
corresponds to the consideration of the case $V/t\rightarrow \infty
,V^{\prime }/t\rightarrow \infty ,V/V^{\prime }=2.$ It allows us to consider
as large clusters as $L=32$. The results of the calculations for the
correlation functions are shown in Figs.3,4. We have found by the technique
described above  $\gamma =2.009\div 2.013$ and $\delta =1.80\div 1.83$.
Note that the envelope of the function $K(R)$ turns out to be nonmonotonous
in non-LL regime (see the black circles for $R=2,4,6$ in Fig.4.

The results of computer simulation demonstrating possible violation of Haldane
hypothesis seem to be rather unexpected. In particular, we cannot see any
simple causes for the difference between two frustrated cases: $\rho=1/2,
V=2V^{\prime}\to\infty$ (non-LL behavior) and $\rho=2/3, V^{\prime}=0,
V\to\infty$ (LL behavior). It would be very important to understand these
numerical results by regular field-theoretical methods.

CAPTIONS\ TO\ FIGURES

Fig.1. The dependence of the ratio $\chi$ (Eq.(1)) on the inverse size of
the cluster; empty symbols correspond to $V'=0$ (circles: $V=0.5t$,
triangles:  $V=1.5t$); black symbols correspond to $V=2V'$ (circles: $V=10t$,
squares:  $V=30t$, triangles: $V=50t$,
diamonds: $V=100t$, hexagons: $V=200t$).

Fig.2. Phase diagram of the model. The boundary between conducting LL and
non-LL phases is shown schematically by zigzags.

Fig.3. The dependence of the correlation functions $G(R)$ (Eq.(5)) for
L=32; open circles correspond to $V=V'=0$, black ones correspond to
$V=2V'$, $V\to\infty$.

Fig.4. The same as in Fig.3, for $K(R)$ (Eq.(5)).

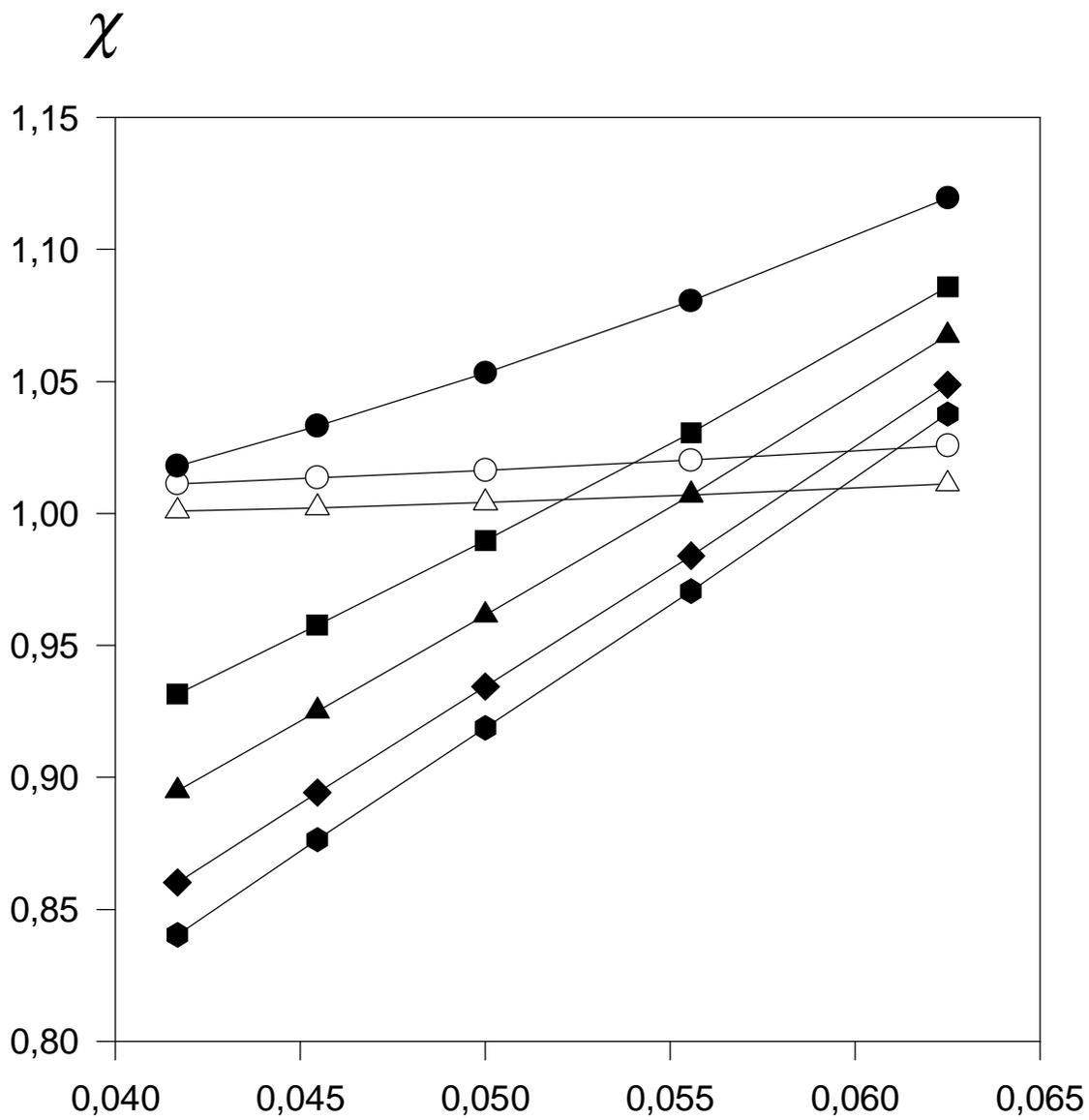
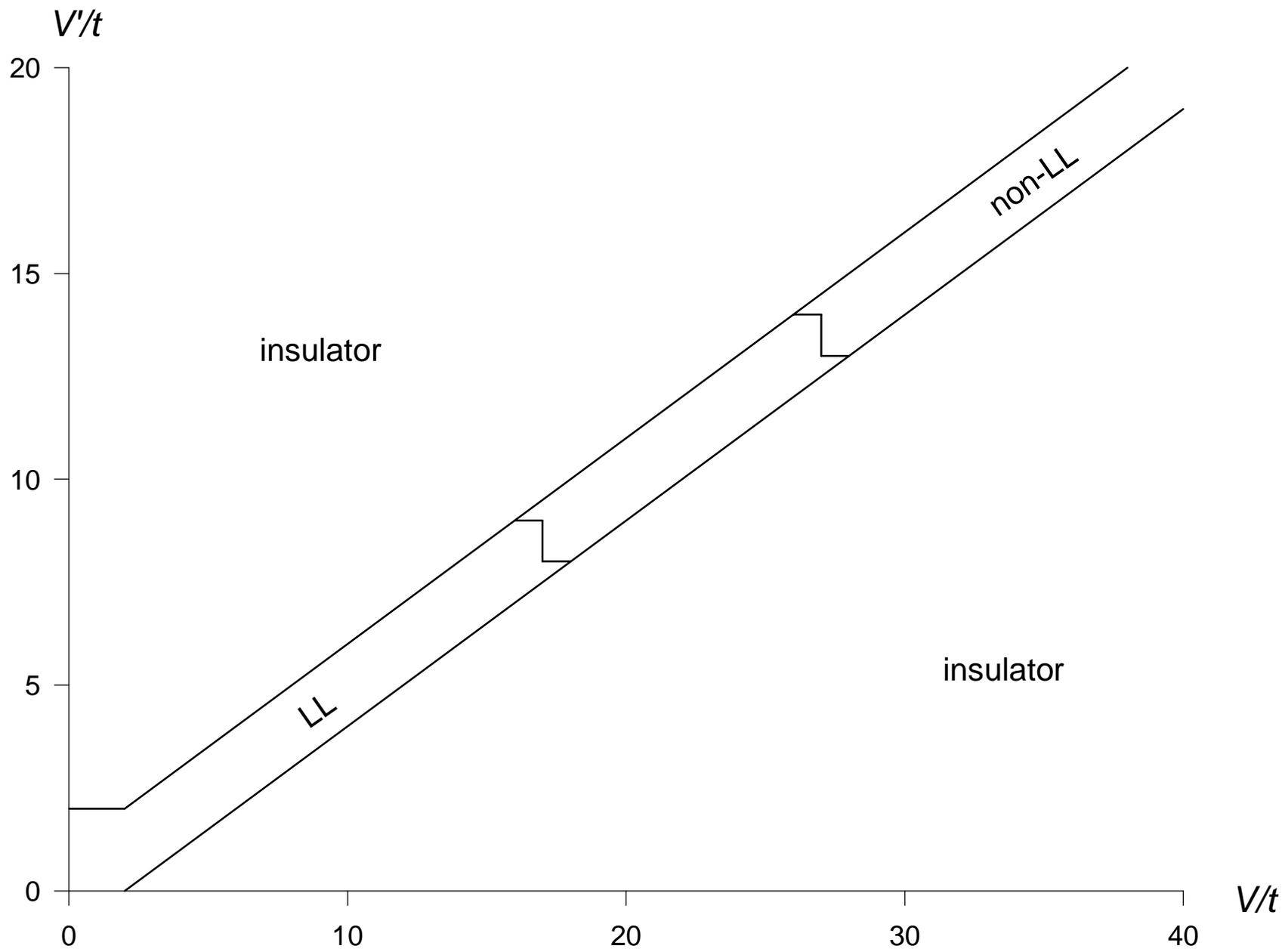
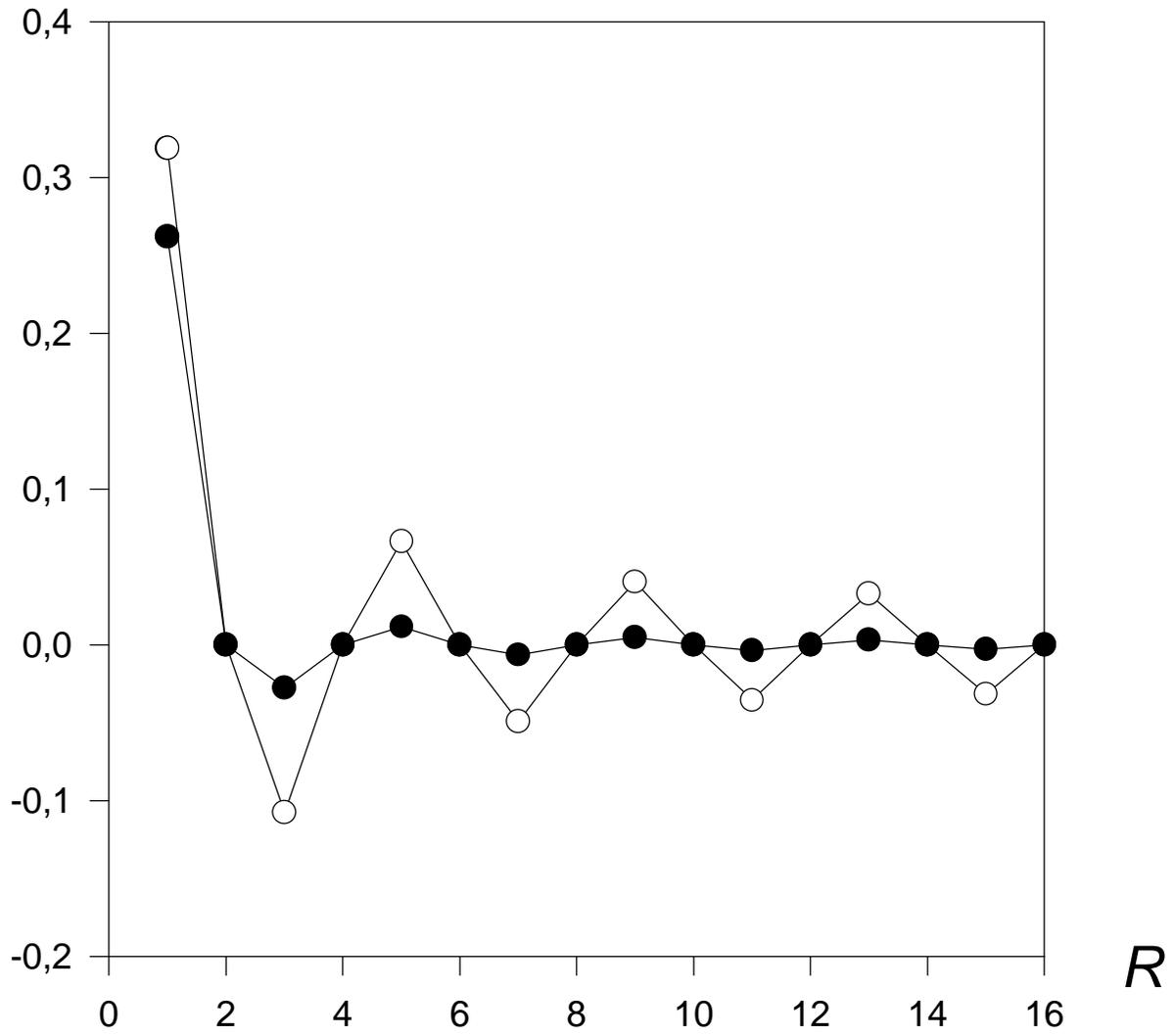
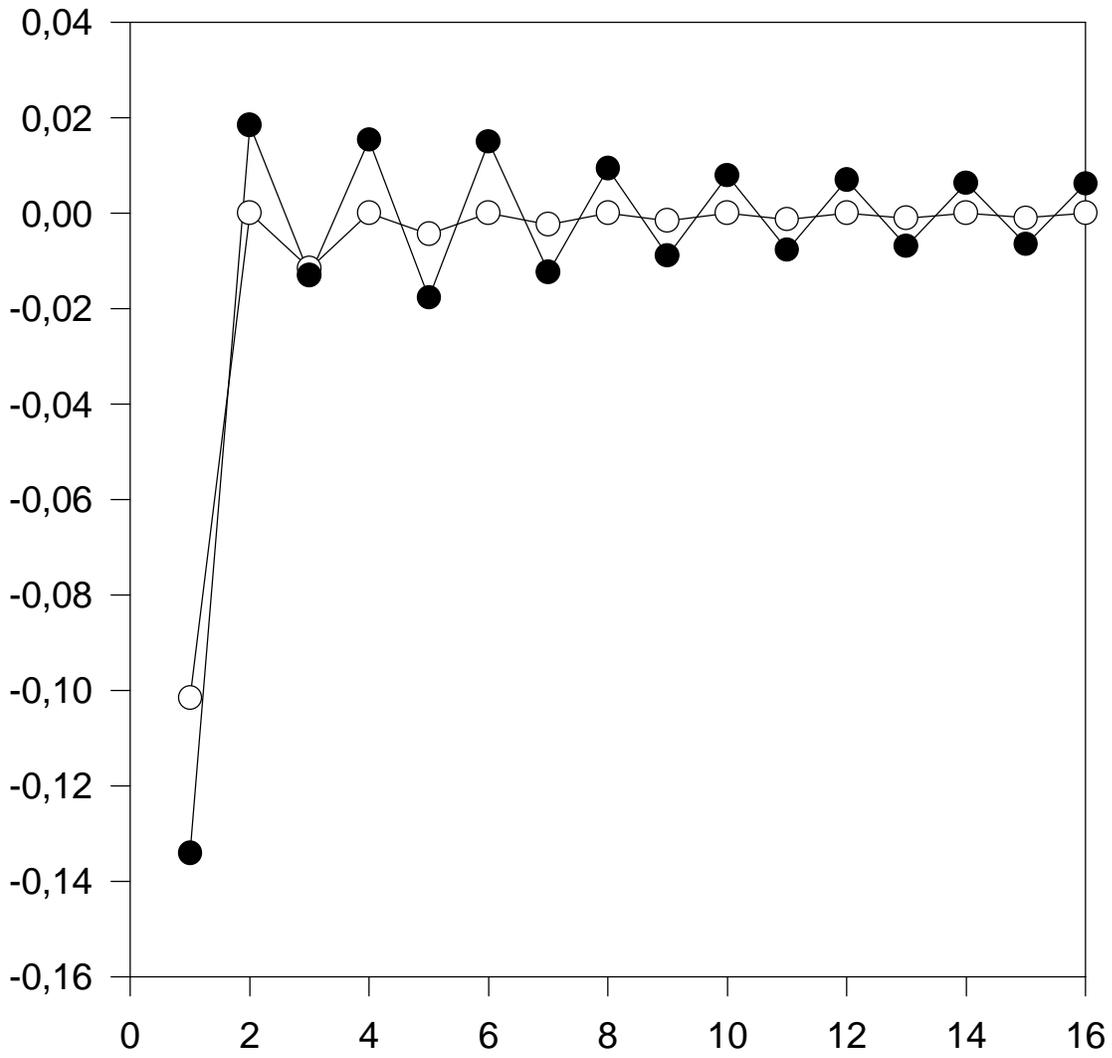
\end{document}